\documentclass[10pt,journal,compsoc]{IEEEtran}

\ifCLASSOPTIONcompsoc
  \usepackage[nocompress]{cite}
\else
  \usepackage{cite}
\fi
\ifCLASSINFOpdf
\else
\fi
\usepackage{graphicx}
\usepackage{algorithm,algorithmic}
\usepackage{amsmath}
\usepackage{amsfonts}
\usepackage{multirow}
\usepackage{wasysym}
\usepackage{color} 
\usepackage{soul}
\newtheorem{theorem}{Theorem}

\begin{document}
%
\title{Energy Efficient Scheduling of Cloud Application Components with Brownout}
%
%
%
%

\author{Minxian~Xu,~\IEEEmembership{}
        Amir Vahid Dastjerdi,~\IEEEmembership{Member, ~IEEE,}
        and~Rajkumar Buyya,~\IEEEmembership{Fellow,~IEEE}
\IEEEcompsocitemizethanks{\IEEEcompsocthanksitem M. Xu, A. V. Dastjerdi and R. Buyya are with Cloud Computing and Distributed Systems (CLOUDS) lab, Department of Computing and Information Systems,
	University of Melbourne, Australia, 3010
	.\protect\\
E-mail: xianecisp@gmail.com
 }
\thanks{Manuscript received： ; revised }}

%
%

\markboth{}%
{}
%



\IEEEtitleabstractindextext{%
\begin{abstract}
It is common for cloud data centers meeting unexpected loads like request bursts, which may lead to overloaded situation and performance degradation. Dynamic Voltage Frequency Scaling and VM consolidation have been proved effective to manage overloads. However, they cannot function when the whole data center is overloaded. Brownout provides a promising direction to avoid overloads through configuring applications to temporarily degrade user experience. Additionally, brownout can also be applied to reduce data center energy consumption. As a complementary option for Dynamic Voltage Frequency Scaling and VM consolidation, our combined brownout approach reduces energy consumption through selectively and dynamically deactivating application optional components, which can also be applied to self-contained microservices. The results show that our approach can save more than 20\% energy consumption and there are trade-offs between energy saving and discount offered to users.
\end{abstract}

}

\maketitle

\IEEEdisplaynontitleabstractindextext

%
\IEEEpeerreviewmaketitle

\ifCLASSOPTIONcompsoc
\IEEEraisesectionheading{\section{Introduction}\label{sec:introduction}}
\else
\section{Introduction}
\label{sec:introduction}
\fi

The emergence of cloud computing is viewed as a new paradigm in IT industry \cite{Buyya}. Cloud computing provides compelling features such as pay-as-you-go pricing model, low operation cost, high scalability and easy access. This makes Cloud computing attractive to business owners as it eliminates the requirement for users to plan ahead for provisioning, and allows enterprises to start with the minimum and request resources on demand. Providers like Amazon, Microsoft, IBM and Google have established data centers to support cloud applications around the world, and aimed to ensure that their services are flexible and suitable for the needs of the end-users.

Energy consumption by the cloud data centers has currently become one of the major problems for the computing industry. The growth and development of complex data-driven applications have promulgated the creation of huge data centers, which heightens the energy consumption \cite{Kaur}. The servers hosted in data centers dissipate more heat and need to be maintained in a fully air-conditioned and engineered environment. The cooling system is already efficient, while servers are still one of the major energy consumer. Hence, reducing server energy consumption has become a main concern of researchers  \cite{Pedram}.

Given the scenario that the budget and resource are limited, overloaded tasks may trigger performance degradation and lead the applications to saturate, in which some applications cannot be allocated by provider. Therefore, some users are not served in a timely manner or experience high latencies, others may even not receive service at all \cite{Klein}. The saturated applications also bring over-utilized situation to hosts and cause high energy consumption. Unfortunately, current resource management approaches like Dynamic Voltage Frequency Scaling (DVFS) and VM consolidation cannot function when the holistic data center is overloaded.  

Currently, applications can be constructed via set of self-contained components that are also called microservices. The components encapsulate its content and expose its functionality through interfaces, which makes them flexible to be deployed and replaced. With components or microservices, developers and users can benefit from their technological heterogeneity, resilience, scaling, ease of deployment, organizational alignment, composability and optimizing for replaceability \cite{Newman}. This brings the advantage of more fine-grained control over the application resource consumption. 

It is common that application components have different priorities to be provided to users. Therefore, not all components or microservices in an application are mandatory to be functional all the time on hosts. We investigate whether it is feasible to downgrade user experience by disabling part of non-mandatory application components or microservices to relieve the over-utilized condition and reduce energy consumption. 

Therefore, we take advantage of a paradigm called \textbf{brownout}. It is inspired by the concept of brownout in electric grids. Its original definition is the voltage shutdown to cope with emergency cases, in which light bulbs emit fewer lights and consumes less power \cite{Durango}. A brownout example for online shopping system is introduced in \cite{Klein}, the online shopping application provides a recommendation engine to recommend similar products that users may be interested in. The recommendation engine component helps service provider to increase the profits, but it is not essential to run the engine. Recommendation engine also requires more resource in comparison to other components. Therefore, with brownout, under overloaded situation, the recommendation engine could be deactivated to serve more clients who require essential requirements. 

There are many other applications with some features can be disabled under brownout situation. Like the online document process application that contains the components for spell checking and report generating. These components are not required to be running all the time and can be deactivated for a while to reduce resource utilization.  Other applications contain components that are not required to be executing all the time can be applied with the brownout approach. What the service providers need to spend efforts on is identifying the optional components and determining their discount when  deactivated. Our motivation is to investigate the trade-off between energy consumption and discount, as well as offering different component selection policies.	

We consider component-level control in our system model. The model could also be applied to container or microservices architecture. We model the application components as mandatory and optional, if required, optional components can be deactivated. By deactivating the optional components selectively and dynamically, the application utilization is reduced and eventually total energy consumption is saved as well. While under market scenario, service provider may provide discount for user as the services are deactivated.

Our objective is to tackle the problem of energy efficiency and our contributions are as below:
\begin{itemize}
	\item Our approach considers the trade-offs between discount that should be given to a user if a component is deactivated and how much of energy can be saved. 
	
	\item Then we propose a number of policies that consider the aforementioned trade-offs and dynamically make decisions on which components are going to be deactivated.

\end{itemize}

The rest of this paper is organized as: after discussing related work in Section 2, we present the brownout enabled system model and problem statement in Section 3. Section 4 introduces our proposed brownout enabled approach in details, while the experimental results of the proposed approach are illustrated in Section 5. The conclusions along with future work are given in Section 6.

\section{Related Work}
It is an essential requirement for Cloud providers to reduce energy consumption, as it can both decrease operating costs and improve system reliability. Data centers can consume from 10 to 100 times more power per square foot than a typical office building. A large body of literature has focused on reducing energy consumption in cloud data centers, and the dominant categories for solving this problem are VM consolidation and Dynamic Voltage Frequency Scaling (DVFS) \cite{Kansal}. 

VM consolidation is regarded to be an act of combining into an integral whole, which helps minimizing the energy consumed by allocating work among fewer machines and turning off unused machines \cite{Pecero}. Under this approach, the VMs hosted on underutilized hosts would be consolidated to other servers and the remaining hosts would be transformed into power-saving state. Beloglazov et al. \cite{ Beloglazov} proposed scheduling algorithms considering Quality of Service and power consumption in Clouds. The algorithm’s objective is energy-efficient mapping VMs
to cloud servers through dynamic VM consolidation. The VM consolidation process is modeled as a bin-packing problem, where VMs are regarded as items and servers are regarded as bins. The advantages of the proposed algorithms are that they are independent of workload types and do not need to know the VM application information in advance.

The authors in \cite{Beloglazov2} introduced adaptive approaches for VM consolidation with live migration according to VM historical data. 
Similar to \cite{Beloglazov}, the VM placement is also modeled as a bin-packing problem, in which VMs from over-utilized servers are allocated to the PM with the least increase of power consumption and under-utilized servers are switched to be off or low power mode. In comparison to \cite{Beloglazov}, this work considers multiple dimension resource (CPU, memory and bandwidth) and focuses more on VM placement optimization stage by proposing various policies. 
This work advances previous work by discussing online algorithm competitive ratio for energy efficient VM consolidation, which proves the algorithm's efficiency. 

A self-adaptive method for VM consolidation on both CPU and memory is introduced in \cite{Mastroianni}. Its objective is minimizing the overall costs caused by energy related issues. The VM assignment and migration processes are determined by probabilistic functions (Bernoulli trial). The mathematical analysis and realistic testbed results show that the proposed algorithm reduces total energy consumption for both CPU-intensive and memory-intensive workloads with suitable Bernoulli trial parameters.  Compared with bin-packing approach (adopted in \cite{Beloglazov} \cite{Beloglazov2}), the proposed algorithm in this work can reduce migration times of VMs and its time complexity is lower than bin-packing-based algorithm, which offers higher efficiency in the online scenario. To achieve the best performance, the disadvantage of this work is that it needs some efforts to find the most suitable Bernoulli trial parameter. Chen et al. \cite{Chen} extended  \cite{Mastroianni} and proposed another utilization-based probabilistic VM consolidation algorithm that aimed to reducing energy consumption and VM migration times. The author also made performance comparison with the algorithms in \cite{ Beloglazov}. 

Corradi et al. \cite{Corradi} considered VM consolidation in a more practical viewpoint related to power, CPU and networking resource sharing and tested VM consolidation in OpenStack, which shows VM consolidation is a feasible solution to reduce energy consumption. Salehi et al. \cite{Salehi} proposed a VM consolidation based approach, which is an adaptive energy management policy that preempts VMs to reduce the energy consumption according to user-specific performance constraints and used fuzzy logic for obtaining appropriate decisions.

Han et al. \cite{Han} used Markov Decision Process (MDP) to handle VM management to reduce data center energy consumption. 
Through MDP, the optimal result is obtained by solving objective function. However, its solution dimension is quite large, the authors also proposed an approximate MDP approach to reduce the solution space and achieve faster convergence. In this approximate algorithm, a centralized controller calculates the utilization function for each VM and determines the possibilities for the state transition. The state transitions in this algorithm represent the VMs are migrated from one server to another.  The authors also theoretically validated the upper bound of algorithm's error. 
 
A practical OpenStack framework is implemented in \cite{Anton2015CCPE} considering  VM consolidation and data center power management.  This framework is available for customized algorithm implementation. With public APIs, the framework is transparent to the base OpenStack installation, and it is not required to modify any OpenStack’s configurations. This work is the first step to implement  VM consolidation in OpenStack to minimize total energy consumption.
 
The DVFS technique introduces a trade-off between computing performance and  energy consumed by the server. The DVFS technique lowers the frequency and voltage when the processor is lightly loaded, and utilizes maximum frequency and voltage when the processor is heavily loaded. Von et al. \cite{Laszewski} introduced a power-aware scheduling algorithm based on DVFS-enabled cluster. 
Hanumaiah \cite{Hanumaiah} et al. introduced a solution that considers DVFS, thread migration and active cooling to control the cores to maximize overall energy efficiency.

The authors in \cite{Kim} modelled real-time service to be real-time VM requests and applied several DVFS algorithms to reduce energy consumption. Their objective is balancing the energy consumption and prices. 
The major concern in this work is that less energy is preferred at the same price, thus three different schemes based on DVFS are proposed to balance the energy consumption and prices. 
The proposed schemes are easy to implement while the adaptive DVFS evaluations are restricted by the simplified and known-in-advance queueing model.

Deng et al. \cite{Deng} proposed a method named CoScale for DVFS coordinating on CPU and memory  while investigating performance constraints, which is the first trial to coordinate them together. Its objective is finding the most efficient frequency from a set of frequency settings while ensuring system performance. The most efficient frequencies for cores and memory are selected as they minimize the whole system energy consumption. 
 CoScale adopts a fine-grained heuristic algorithm that iteratively predicates the component frequencies according to its performance counters as well as online models. 
However, CoScale is not suitable for offline workloads because it cannot reduce the possible frequency space as like in online workloads.


Teng et al. \cite{Teng} combined DVFS and VM consolidation together to minimize total energy consumption. 
The energy saving objective is mainly applied to batch-oriented scenario, in which the authors introduced a DVFS-based algorithm to consolidate VMs on servers to minimize energy consumption and ensure job Service Level Agreement. 
With theoretical analysis and realistic testbed on Hadoop, the authors proved that the proposed algorithm can find the most efficient frequency that is only associated with the processor type and its VM consolidation performance is insensitive to tunable parameters.  The limitations of this work 
is that its realistic testbed is already upgraded to a new version that provides better management, which is more persuasive to implement the proposed approach on the updated platform.

Brownout was originally applied to prevent blackouts through voltage drops in case of emergency. Klein et al. \cite{Klein} firstly borrowed the approach of brownout and applied it to cloud applications, aiming to design more robust applications under unpredictable loads. Tomas et al. \cite{Tomas} used the brownout along with overbooking to ensure graceful degradation during load spikes and avoid overload. Durango et al. \cite{Durango} introduced novel load balancing strategies for applications by supporting brownout. In a brownout-compliant application or service, the optional parts are identified by developers and a control knob called \textbf{dimmer} that controls these optional parts is also exported. The dimmer value represents a certain probability given by a control variable and shows how often these optional parts are executed. In addition, a brownout controller is also required to adjust the dimmer value to avoid overload \cite{Tomas}.  

To the best of our knowledge, our approach is the first research to reduce energy consumption with brownout at components level, which also considers  revenues for cloud service providers. Our approach provides a complementary option apart from VM consolidation and DVFS.

\section{Problem Statement}

 In this section, we explain our system model and state the problem we aim to tackle. For reference, Table 1 summaries the symbols and their definitions throughout this paper.


 \subsection{System Model}
  Our system model (Fig. 1) includes entities: users, applications and cloud providers, which are discussed as below: 
 
 \textbf{Users}:  Users submit service requests to cloud data centers to process. User entities contain user id and requested applications (services) information. 
 
 \textbf{Applications}: The application entities in our model come into system together with user entities. The applications consist of a number of components, which are tagged as mandatory or optional. 
  
 \textbf{Mandatory component}: The mandatory component is always running (activated) when the application is executed. 
 
 \textbf{Optional component}: The optional component can be set as activated or deactivated. These components have parameters like \textit{utilization} and \textit{discount} (penalty payment amount). Utilization indicates the amount of reduced utilization, and discount represents the price that is cut. The deactivation and activation of optional components are controlled by the \textbf{brownout controller}, which makes decisions based on system status and component selection policies. 

The components can also be \textit{connected}, which means that they  communicate with each other and there are data dependencies between them. Therefore, we consider that if a component is deactivated, then all its connected optional components would also be set as deactivated. For example in Fig. 1, if Com3 in Application \#1 is deactivated, Com2 should also be deactivated; in Application \#2, if Com1 is deactivated, Com3 should also be deactivated; in Application \#n, if Com4 is deactivated, Com3 should also be deactivated, but Com2 is still working (Com1 is connected with Com3, but Com1 is mandatory, so it is not deactivated).
 
 \textbf{Cloud Providers}: Cloud providers offer resources to meet service demands, which host a set of VMs or containers to run applications.

 \begin{table}[]

 	\centering
 	\caption{Symbols and Definitions}
 	\label{my-label}
 	\resizebox{0.48\textwidth}{!}{%
 		\begin{tabular}{|c|l|}
 			\hline
 			\textbf{Symbol}           & \multicolumn{1}{c|}{\textbf{Definition}}                 \\ \hline
 			$h_i$                     & Server (host) $i$                                        \\ \hline
 			$P_i^{server}$            & Power of  $h_i$                             \\ \hline
 			$P_i^{idle}$              & Power when  $h_i$ is idle                                \\ \hline
 			$P_i^{dynamic}$           & Power when  $h_i$ is fully loaded                        \\ \hline
 			$P_i^{max}$               & Maximum power of  $h_i$                                  \\ \hline
 			$hl$                      & Server list in data center                               \\ \hline
 			$w_i$                     & Number of VMs assigned to  $h_i$                         \\ \hline
 			$VM_{i, j}$               & VM $j$ on  $h_i$                                         \\ \hline
 			$u(VM_{i, j})$            & Utilization of VM $j$ on  $h_i$                          \\ \hline
 			$d(VM_{i, j})$            & Discount of VM $j$ on  $h_i$                             \\ \hline
 			$App_c$                   & Application component $c$                                \\ \hline
 			$A_j$                     & Total number of application components                   \\ \hline
 			$u(App_c)$                & Utilization of application component $c$                 \\ \hline
 			$d(App_c)$                & Discount of application component $c$                    \\ \hline
 			$D_i$                     & Total discount from server $i$                           \\ \hline
 			$N$                       & Total number of VMs                                      \\ \hline
 			$M$                       & Total number of servers                                  \\ \hline
 			$Eff_{pa}$                & Algorithm efficiency of proposed algorithm $pa$          \\ \hline
 			$E_{pa}$                  & Energy consumption of proposed algorithm $pa$            \\ \hline
 			$E_{bl}$                  & Energy consumption of baseline algorithm $bl$            \\ \hline
 			$D_{pa}$                  & Discount amount of proposed algorithm $pa$                      \\ \hline
 			$\alpha$                  & Weight of discount to calculate algorithm efficiency     \\ \hline
 			$t$                       & Time interval $t$                                        \\ \hline
 		\color{black}	$T$                       & \color{black} The whole scheduling interval \\ \hline
 			$TP$                      & Overloaded power threshold                               \\ \hline
 			$\theta_t$                & Dimmer value in brownout at time $t$                     \\ \hline
 			$n_t$                     & Number of overloaded hosts at time $t$                   \\ \hline
 			$P_i^{r}$           & Expected power reduction of $h_i$                                   \\ \hline
 		\color{black}	$COH()$               & \color{black}  Calculate the number of overloaded hosts           \\ \hline
 		\color{black}	$HPM()$                  & \color{black}Host power model to compute expected utilization reduction                    \\ \hline
 			$u_{h_i}^{r}$       & Expected utilization reduction on $hi$                   \\ \hline
 		\color{black}	$VUM()$                   & \color{black}VM utilization model to compute expected utilization reduction                       \\ \hline
 			$u_{VM_{i, j}}^{r}$ & Expected utilization reduction on $VM_{j}$ on $h_i$                      \\ \hline
 			$CSP()$                     & Component selection policy to deactivate components \\ \hline
 			$dcl_{i, j,t}$              & Deactivated component list at time $t$ on $h_i$          \\ \hline
 			$S_t$                     & Set of deactivated components connection tags                 \\ \hline
 			$Ct(App_c)$               & Connection tag of component $App_c$                           \\ \hline
 			$ocl_{i, j, t}$            & Optional component list of $VM_{j}$ on $h_i$ at time $t$   \\ \hline
 			$p$                       & Position index in optional component list                     \\ \hline	
 		\color{black}	$R_h$                    & \color{black}Available resource of host \\ \hline
 		\color{black}	$R_v$        & \color{black}Maximum requested resource of VM \\ \hline
 		\color{black}	$C_e$        &\color{black} Cost of energy consumption per unit of time \\ \hline
 		\color{black}	$C_o$        & \color{black}Cost of overloads per unit of time \\ \hline
 		\color{black}	$\varepsilon$ & \color{black}Relative cost of overloads compared with $C_e$\\ \hline
 		\color{black}	$t_b$ &\color{black} Time for brownout operation \\ \hline
 		\color{black}	$t_m$ &\color{black} Time for VM consolidation \\ \hline
 		\color{black}	$\tau$  &\color{black} The times of brownout and VM consolidation occur in $T$\\ \hline
 		\end{tabular}%
 	}
 \end{table}
 
  \begin{figure}[ht]
  	\centering
  	\includegraphics[width=0.9\linewidth]{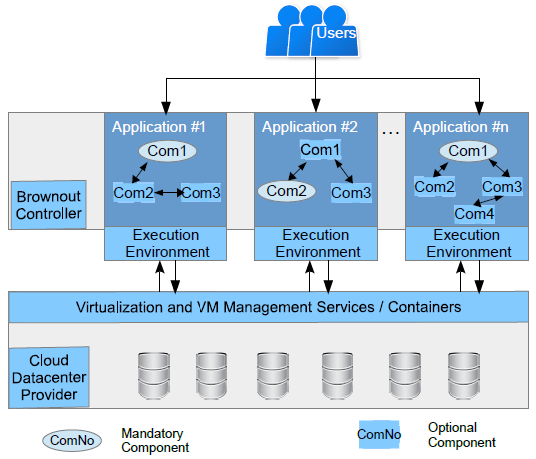}
  	\caption[Brownout Architecure]{System Model with Brownout}
  	
  	\label{fig:BrownoutArchi}
  \end{figure}
 
\subsection{Power Model}

 To calculate the total energy consumption of data center, we adopt the server power model proposed by Zheng et al. \cite{Zheng}.
 The server power consumption is modeled as:
   
{\small { \begin{equation}
	P_i^{server} = 	\begin{cases}
		P_i^{idle} + \sum_{j=1}^{w_i} u(VM_{i,j}) \times P_i^{dynamic}   & ,w_i > 0\\
		0 & ,w_i = 0
	\end{cases}
\end{equation}}}
$P_i^{server}$ is composed of idle power and dynamic power. The idle power is regarded as constant and the dynamic power is linear to the total CPU utilization of all the VMs on the server. If no VM is hosted on a server,  the server is turned off to save power. $VM_{i,j}$ refers to the $j$th VM on the $i$th server, $w_i$ means the number of VMs assigned to server $i$.
 
{\small \begin{equation}
u(VM_{i, j}) = \sum_{c=1}^{A_j} u(App_c)
\end{equation} 

 The utilization of $VM_{i, j}$ is represented as $u(VM_{i,j})$, which is computed by summing up all the application utilization on the $j$th VM. The $c$ is the component id and $A_j$ is the number of application components. As processors are still the main energy consumer of servers, we focus on CPU utilization in this work.
 
 \subsection{Discount Amount}
 
 \begin{equation}
D_i = \sum_{j=1}^{w_i}d(VM_{i,j})
 \end{equation} 
 
  In equation (3), $D_i$ is the total discount amount obtained from all VMs, in which the individual discount $d(VM_{i,j})$ from $VM_{i,j}$ is the sum of all application components discount amount $d(App_c)$ as shown in equation (4).
 
 \begin{equation}
 d(VM_{i,j}) = \sum_{c=1}^{A_j} d(App_c)
 \end{equation} }
$A_j$ is the number of applications hosted on $VM_j$, and  $d(VM_{i,j})$  is the discount happened from $VM_j$ on server $i$, and $D_i$ is the total discount amount on server $i$. 

\subsection{Constraints and Objectives}
 The above equations subject to the following constraints:
 
{\small   \begin{equation}
  \sum_{i=1}^{M}w_i = N
  \end{equation} }
  
{\small     \begin{equation}
    \sum_{j=1}^{w_i} u(VM_{i,j}) \leq 1, \forall i \in [1, M]
    \end{equation} }
 
 $N$ is the total number of VMs and $M$ is the total number of servers. Equation (5) represents the total number of VMs assigned to hosts $w_i$ equals to the sum of VMs. Equation (6) represents the sum of all the VMs utilization cannot surpass their host available utilization. 
 
 We formulate the objectives of this problem as:
 {\small     \begin{equation}
 	\min{\sum_{i=1}^{M} P_i^{server}}
 	\end{equation} }
 
 As well as
 {\small     \begin{equation}
 \min(\sum_{i=1}^{M} D_i)
 	\end{equation} }
 Therefore, we aim at investigating the trade-off total energy consumption and discount amount. 
 
 To measure the performance of an algorithm, we represent the algorithm efficiency $Eff_{pa}$:  
\begin{equation}
Eff_{pa} = \frac{E_{pa}}{E_b} +  \alpha D_{pa}
\end{equation}
where $E_{pa}$ is the energy consumption of the proposed algorithm, $E_b$ is the baseline algorithm energy  consumption, $D_{pa}$ is the discount amount offered by the proposed algorithm. If the proposed algorithm saves more energy than the baseline algorithm, $\frac{E_{pa}}{E_b}$ is a value between 0 to 1, and $D_{pa}$ represents the offered discount percentage, which also belongs to 0 to 1. Thus, the smaller $Eff_{pa}$ is, the more energy is reduced and less discount amount is offered.
The $\alpha$ is the weight of discount, its default value is 1.0, if service provider care more on discount, $\alpha$ is set as larger than 1.0; if they care more about energy saving, $\alpha$ is set as less than 1.0.

\section{Proposed Approach}

 \begin{algorithm}
 	\color{black}
 	\footnotesize
 	\caption{Energy Efficient with Brownout Algorithm (EEBA)}
 	\begin{algorithmic}[1]
 		
 		\renewcommand{\algorithmicrequire}{\textbf{Input:}}
 		\renewcommand{\algorithmicensure}{\textbf{Output:} }
 		\REQUIRE  hostList $hl$ with size $M$, application components information, overloaded power threshold $TP$, dimmer value $\theta_t$ at time $t$, scheduling interval $T$, deactivated component list $dcl_{i, j, t}$ of $VM_{i,j}$ on host $h_i$, power model of host $HPM$, VM utilization model $VUM$, component selection policy $CSP$
 		\ENSURE   total energy consumption, discount amount, number of shutting down hosts 
 		\\ 
 		\STATE use PCO algorithm to initialize VMs placement
 		\STATE initialize parameters with inputs, like $TP$  

 		\FOR {$t \leftarrow 0$ to $T$}
 		
 		\STATE $n_t \leftarrow COH(hl)$
 		\IF{$n_t > 0$}
 		\STATE  $\theta_t$ $\leftarrow$ = $ \sqrt{\frac{n_t}{M}}$
 		
 		\FORALL {$h_i$ in $hl$  (i.e. $i = 1, 2, \dots, M$) }
 		\IF{($P_{i}^{server}$ $\textgreater$ $P_{i}^{max}$ $\times$ $ TP$)}
 		\STATE $P^{r}_i$ $\leftarrow$ $\theta_t$ $\times$ $P_{h_i}$
 		\STATE $u_{h_i}^{r}$ $\leftarrow$ $HPM$($h_i, P^{r}_i$)
 		\FORALL {$ VM_{i,j} $ on $h_i$ (i.e. $j = 1, 2, \dots, w_i$)}
 		 		\STATE $dcl_{i, j, t}$ $\leftarrow$ NULL
 		\STATE $u_{VM_{i, j}}^{r}$ $\leftarrow$ $VUM$($u_{h_i}^{r}$, $VM_{i,j}$)
 		\STATE $dcl_{i, j, t}$ $\leftarrow$ $CSP$($u_{VM_{i, j}}^{r}$)
 		\STATE $D_i \leftarrow D_i + d(VM_{i, j})$
 		\ENDFOR
 		\ENDIF
 		\ENDFOR
 		\ELSE 
 		\STATE activate deactivated components
 		\ENDIF

 		\STATE use VM consolidation in PCO algorithm to optimize VM placement
 		\ENDFOR

 	\end{algorithmic} 
 \end{algorithm}
	
Prior to brownout approach, we require a VM placement and consolidation algorithms. We adopt the placement and consolidation algorithm (PCO) proposed by Beloglazov et al. \cite{ Beloglazov }.　 Then we propose our brownout enabled algorithm based on PCO and introduce a number of component selection policies considering component utilization and discount.

\subsection{VM　 Placement and Consolidation Algorithm (PCO)}

 The VM placement and consolidation (PCO) algorithm is an adaptive heuristics for dynamic consolidation of VMs and extensive experiments show that it can significantly reduce energy consumption. In the initial VM placement phase, PCO sorts all the VMs in decreasing order of their current CPU utilization and allocates each VM to the host that increases the least power consumption due to this allocation. In the VM consolidation phase, PCO optimizes VM placement according to loads of hosts: PCO separately picks VMs from over-utilized and under-utilized hosts to migrate, and finds new placements for them. After migration, the over-utilized hosts are not overloaded any more and the under-utilized hosts are switched to sleep mode. 

\subsection{Energy Efficient Brownout Enabled Algorithm}

 \color{black} Our proposed energy efficient brownout enabled approach (noted as \textbf{EEBA}) is an enhanced approach based on PCO algorithm. According to host power consumption, the brownout controller dynamically deactivates or activates applications' optional components on VMs to relieve overloads and reduce the power consumption. 

 As shown in Algorithm 1, EEBA mainly consists of 6 steps:

Before entering the approach procedures, service provider firstly needs to initialize VM placement by algorithm like PCO and overloaded power threshold (lines 1-2). The power threshold $TP$ is a value for checking whether a host is overloaded. Then the other steps are as below:  

1) In each time interval $t$, checking all the hosts and counting the number of  overloaded hosts as $n_t$ (line 4);

2) Adjusting dimmer value $\theta_t$ as $\sqrt{ \frac{n_t} {M}}$ based on the number of overloaded hosts $n_t$ and host size $M$ (line 6). 

As mentioned in our related work, the dimmer value $\theta_t$ is a control knob used to determine the adjustment degree of power consumption at time $t$. The dimmer value $\theta_t$ is 1.0 if all the hosts are overloaded at time $t$ and it means that brownout controls components on all the hosts. The dimmer value is 0.0 if no host is overloaded and brownout will not be triggered at time $t$. The dimmer adjustment approach shows that dimmer value varies along with the number of overloaded hosts. 

3) Calculating the expected utilization reduction on the overloaded hosts (lines 8-10). According to the dimmer value and host power model, EEBA calculates expected host power reduction $P_i^r$ (line 9) and expected utilization reduction $u_{h_i}^r$ (line 10). With host power model (like in Table 3), we have host power at different utilization levels, so the utilization reduction can be computed based on power reduction. For example, in a power model, the host with 100\% utilization is 247 Watts and 80\% utilization is 211 Watts, if the power is required to be reduced from 247 to 211 Watts, the expected utilization reduction is  $100\%-80\% = 20\%$.    

4) Calculating the expected utilization reduction on VM (lines 11-13). An empty deactivated component list $dcl_{i, j, t}$ of $VM_{j}$ on host $h_i$ is initialized to prepare for storing deactivated components (line 12).  Then the expected VM utilization reduction $u_{VM_{i,j}}^r$ is computed based on VM utilization model as VM utilization multiplies $u_{h_i}^r$ (line 13).

5) Applying component selection policy $CSP$ to find and deactivate components list $dcl_{i, j,t}$ (line 14). According to the expected VM utilization reduction $u_{VM_{i,j}}^r$, component selection policy is responsible for finding the components satisfying the utilization constraint, deactivating these components and their connected ones, and updating total discount amount (line 15).

6) In EEBA, if no host is above the power threshold, the algorithm activates the optional components that have been set as deactivated (line 20).

Finally, after finishing the main steps of EEBA, VM consolidation in PCO algorithm is applied to optimize VM placement (line 22). 

\color{black} The EEBA algorithm takes effect between the VM placement and VM consolidation in PCO. VMs are initially placed by VM placement phase in PCO, after that, if no host is above the power threshold, the EEBA does not work; otherwise, the brownout is triggered to handle the overloaded condition, then the VM consolidation phase in PCO is applied. 


 \begin{algorithm}
 
 	\footnotesize 
 	\caption{Component Selection Policy: Lowest Utilization First Component Selection Policy (LUFCS)}
 	\begin{algorithmic}[1]
 		 
 		\renewcommand{\algorithmicrequire}{\textbf{Input:}}
 		\renewcommand{\algorithmicensure}{\textbf{Output:} }
 		\REQUIRE expected utilization reduction $u_{VM_{i,j}}^r$ on $VM_{i, j}$
 		\ENSURE  deactivated components list $dcl_{i,j,t}$ 
 		\\ 
 		\STATE Sort the optional component list $ocl_{i,j, t}$ based on utilization $u(App_c)$ in ascending order //Other policies may change the sorting approach at this line. If there are connected components, the connected components are treated together and sorted by their average utilization
 		
 		\STATE  $S_{t}$$\leftarrow$ NULL

 		\IF{$u(App_{1st})$ $\geq$ $u_{VM_{i, j}}^{r}$}
 		\STATE  $dcl_{i, j, t}$ $\leftarrow$  $dcl_{i,j, t}$ + $App_{1st}$
 		\STATE  $S_{t}$$\leftarrow$ $S_{t}$ + $Ct(App_{1st})$
 		\FORALL {$App_c$ in $ocl_{i,j, t}$}
 		
 		\IF{$Ct(App_c)$ is in $S_t$}
 		\STATE $dcl_{i, j, t} \leftarrow dcl_{i, j, t} + App_c$
 		\STATE $ d(VM_{i, j})  \leftarrow  d(VM_{i, j})  + d(App_c)$
 		
 		\ENDIF
 		\ENDFOR

 		\ELSE 
 		\STATE　 $p \leftarrow 0$
 		
 		\FOR {$App_c$ in $ocl_{i,j, t}$}

 		\IF {($\sum_{0}^{k} (App_c)$ $< u_{VM_{i, j}}^{r}$ \& $\sum_{0}^{k+1} (App_c)$ $> u_{VM_{i, j}}^{r}$)}
 		\color{black} \IF {($ u_{VM_{i, j}}^{r} - \sum_{0}^{k} (App_c) < \sum_{0}^{k + 1} (App_c) - u_{VM_{i, j}}^{r}  $)} 
 		 \STATE $p = k - 1$
	\ELSE 
	\STATE  $p = k$
 	\ENDIF		
	\STATE \textbf{break} 
 	\ENDIF
 		\ENDFOR
 		
 		\FOR {$c \leftarrow $$0$ to $p$}
 		\STATE  $dcl_{i, j, t} \leftarrow dcl_{i, j, t} + App_c$
 		\STATE  $S_{t}$$\leftarrow$ $S_{t}$ + $Ct(App_{c})$
 		\STATE　$ d(VM_{i, j})  \leftarrow  d(VM_{i, j})  + d(App_c)$
 		
 		\ENDFOR
 		
 		\FORALL {$App_c$ in $ocl_{i, j, t}$}
 		 		
  		\IF{$Ct(App_c)$ in $S_t$}
 		\STATE $dcl_{i, j, t} \leftarrow dcl_{i, j, t} + App_c$
 		\STATE $d(VM_{i, j}) \leftarrow d(VM_{i, j}) + d(App_c)$
 		\ENDIF
 		\ENDFOR

 		\ENDIF
 		\STATE　\textbf{return} $dcl(i, j, t)$

 	\end{algorithmic} 
 \end{algorithm}
\color{black}

 As applications may have multiple optional components with different utilization and discount amount, for Algorithm 1 step 4 that applies component selection policy, we have designed several policies:
 
\textbf{Nearest Utilization First Component Selection Policy (NUFCS)}: The objective of NUFCS is finding and deactivating a single component in the component list. Compared with other components, the single component has the nearest utilization to $u_{VM_{i,j}}^r$. NUFCS can find the goal component in $O(n)$ time, which is efficient in online scheduling.

If the deactivated component is connected with other components, NUFCS also deactivates other connected components. NUFCS runs fast and can reduce utilization, but if $u_{VM_{i,j}}^r$ is much larger than all the single component utilization in the component list, more components should be deactivated to achieve expected energy reduction. Therefore, we propose another 3 multiple components selection policies to achieve expected utilization reduction: 

\textbf{Lowest Utilization First Component Selection Policy (LUFCS)}: LUFCS selects a set of components from the component with the lowest utilization until these components achieve expected utilization reduction. This policy follows the assumption that the component with less utilization is less important for users. Hence, with this policy, the service provider deactivates a number of components with low utilization to satisfy the expected utilization reduction. 

\textbf{Lowest Price First Component Selection Policy (LPFCS)}: LPFCS selects a set of components from the component with lowest discount. This policy focuses more on discount and its objective is deactivating a number of components with less discount amount and satisfying the expected utilization reduction.  

\textbf{Highest Utilization and Price Ratio First Component Selection Policy (HUPRFCS)}: HUPRFCS selects a set of components considering component utilization and discount together. The components with larger $\frac{u(App_c)}{d(App_c)}$ values are prior to be selected. Its objective is deactivating the components with higher utilization and smaller discount. Therefore, the service provider saves more energy while offering less discount amount.

Algorithm 2 shows an example about how the component selection policy works. The example is about LUFCS: the input of the algorithm is the expected configured utilization $u_{VM_{i,j}}^r$ and the output is the deactivated components list $dcl_{i, j, t}$. The steps are:

a) Algorithm 2 sorts the optional components list $ocl_{i ,j ,t}$ based on component utilization parameter in ascending sequence (line 1), therefore, the component with the lowest utilization is put at the head. For connected components, the sorting process is modified as treating the connected components together and using their average utilization for sorting, which lowers the priority of deactivating connected components to avoid deactivating too many components due to connections; 

b) Initialize a set $S_t$ that stores the deactivated components connection tags (line 2);

c) Algorithm 2 deactivates the first component and its connected components if it satisfies the expected utilization reduction (lines 3-11). If the first component utilization parameter value is above $u_{VM_{i,j}}^r$, Algorithm 2 puts this component into the deactivated components list $dcl_{i,j,t}$ and puts its connection tag $Ct(App_1st)$ 
(a tag shows how it is connected with other components) 
into $S_t$. After that, Algorithm 2 finds other connected components and put them into deactivated components list. Finally, summing up the deactivated components discount amount as $d(VM_{i, j})$; 

d) If the first component utilization does not satisfy the expected utilization reduction, Algorithm 2 finds a position \textit{p} in the optional components list (lines 13-23). The sublist before $p-1$ is the last sublist that makes its components utilization sum less than $u_{VM_{i,j}}^r$ and the sublist that before $p$ is the first sublist that makes its components utilization sum larger than $u$. The policy selects the sublist with utilization sum closer to the $u_{VM_{i,j}}^r$ from these two sublists; 

e) Algorithm 2 puts all the components in the sublist into the deactivated components list and puts their connection parameters into the $S_t$ (lines 24-28);  

f) Algorithm 2 finds other connected components and puts them into the deactivated components list, and updates the discount amount (lines 29-35); 

g) Finally, Algorithm 2 returns the deactivated components list (line 36).

The LPFCS and HUPRFCS procedures are quite similar to LUFCS except the sorting process at line 1. For example, the LPFCS sorts the optional components list according to component discount, while HUPRFCS sorts the optional components list based on component utilization and discount ratio $\frac{u(App_c)}{d(App_c)}$. For connected components, these policies also treat them together and use their discount or utilization and discount ratio to sort.

 The complexity of our proposed algorithm at each time interval is calculated based on two parts, one is the brownout part and the other is the PCO part. At each time interval, the complexity of the brownout part is $O(m*M) $, where $m$ is the maximum number of components in all applications, $M$ is the number of hosts. The complexity of the PCO part is $O(2M) $ as analyzed in \cite{Beloglazov}. The complexity at each time interval of our proposed algorithm is the sum of the two parts, which is to $O((2+m)*M) $.
 
 \color{black}
 \subsection{EEBA Competitive Analysis} 
 We apply competitive analysis \cite{Borodin}\cite{Beloglazov2} to analyze the brownout approach combining with VM consolidation for multiple hosts and VMs. 
We assume that there are $M$ homogeneous hosts and $N$ homogeneous VMs. If the available resource of each host is $R_h$ and maximum resource that can be allocated to VM is $R_v$, then the maximum number of VMs allocated to host is $\frac{R_h}{R_v}$.
Overloaded situation occurs when VMs require more capacity than $R_h$. The brownout approach handles with the overloaded situation with a processing time $t_b$, and VMs are migrated between hosts through VM consolidation with migration time $t_m$. The cost of overloads per unit of time is $C_o$, and the cost of energy consumption is $C_e$.
Without loss of generality, we can define $C_e = 1$ and $C_o = \varepsilon$.
Then we have the following theorem: 
\begin{theorem}	
The upper bound of the competitive ratio of EEBA algorithm for the components control and VM migration problem is $\frac{EEBA(I)}{OPT(I)} \leq 1 + \frac{N\varepsilon}{N+M}$.
\end{theorem}

 \begin{IEEEproof}
The EEBA controls the application components to handle with the overloaded situation and applies VM consolidation to reduce energy consumption. This algorithm deactivates application components to make the hosts to be not overloaded and consolidates VMs to the minimum number of hosts. 
Under normal status, the number of VMs allocated to each host is $\frac{N}{M}$, while in overloaded situation, at least $\frac{N}{M}+1$ VMs are allocated to a single host. Thus, the maximum number of overloaded hosts is $M_o = \lfloor\frac{N}{\frac{N}{M}+1}\rfloor$, which is  equivalent to $M_o = \lfloor\frac{MN}{N+M}\rfloor$.

In the whole scheduling interval $T$, we split the time into 3 parts $T = (t_b + t_m)\tau + t_0$, where $t_b$ is the time that EEBA uses brownout to relieve overloads, $t_m$ is the time consumed for VM migration, $t_0$ is the time that hosts running at normal status and $\tau$ $\in \mathbb{R^+}$ . For the brownout and VM migration parts, the behaviors are as below:

1). During the $t_b$, the brownout controller selects application components on overloaded hosts and deactivates them. Because all the hosts are active during $t_b$, the cost of this part is $t_b(MC_e + M_oC_o)$.

2). During the  $t_m$, if there are still overloaded hosts, VMs are migrated from the overloaded hosts $M_{o}^{'}$, and $M_{o}^{'} \leq M_o$. As the VM migration time is $t_m$ and all the hosts are active during migration, the total cost during this time of period is $t_m(MC_e+ M_{o}^{'}C_o)$.

Therefore, the total cost $C$ during $t_b + t_m$ is defined as below:
\begin{equation}
C = t_b(MC_e + M_oC_o) + t_m(MC_e+ M_{o}^{'}C_o).
\end{equation}

And the total cost incurred by EEBA for the input $I$ is shown in equation (11):
\begin{equation}
EEBA(I) = \tau [t_b(MC_e + M_oC_o) + t_m(MC_e+ M_{o}^{'}C_o)]
\end{equation}

The optimal offline algorithm for this problem only keeps the VMs at each host and does not apply brownout and VM consolidation. Therefore, the total cost of an optimal offline algorithm is defined as:
\begin{equation}
OPT(I) = \tau(t_b + t_m)MC_e
\end{equation}

Then we compute the competitive ratio of an optimal offline deterministic algorithm as:
\begin{equation}
\frac{EEBA(I)}{OPT(I)} = \frac{\tau [t_b(MC_e + M_oC_o) + t_m(MC_e+ M_{o}^{'}C_o)]}{\tau(t_b + t_m)MC_e}
\end{equation}
As $M_{o}^{'} \leq M_o$, we have:
\begin{equation}
\frac{EEBA(I)}{OPT(I)} \leq \frac{\tau [(t_b+t_m)(MC_e + M_oC_o)]}{\tau(t_b + t_m)MC_e} = \frac{MC_e + M_oC_o}{MC_e}
\end{equation}
As $M_o = \lfloor\frac{MN}{N+M}\rfloor$, we have $M_0 \leq \frac{MN}{N+M}$ and combine with equation (14) as well $C_e =1, C_o = \varepsilon$, the competitive ratio is defined as:
\begin{equation}
\frac{EEBA(I)}{OPT(I)} \leq \frac{MC_e + M_oC_o}{MC_e} \leq \frac{M + \frac{MN}{N+M}\varepsilon}{M} = 1+ \frac{N\varepsilon}{N+M}
\end{equation}
\end{IEEEproof}

\color{black}

\section{Performance Evaluation}
\subsection{Environment Setting}

\begin{table}
	\centering
		 \caption{Host / VM Types and Capacity}
	\resizebox{0.48\textwidth}{!}{%

\begin{tabular}{|l|l|l|l|l|l|}

	\hline  Name & CPU & Cores  &  Memory & Bandwidth & Storage \\ 
	\hline
	\hline Host Type 1 & 1.86 GHz & 2 & 4 GB & 1 Gbit/s & 100 GB \\ 
	\hline Host Type 2 & 2.66 GHz & 2 & 4 GB & 1 Gbit/s & 100 GB  \\
			\hline
			\hline VM Type 1 & 2.5 GHz & 1 & 870 MB & 100 Mbit/s & 1 GB \\ 
			\hline VM　Type 2 & 2.0 GHz & 1 & 1740 MB & 100 Mbit/s & 1 GB  \\
			\hline VM Type 3 & 1.0 GHz & 1 & 1740 MB & 100 Mbit/s & 1 GB  \\
			\hline VM Type 4 & 0.5 GHz & 1 & 613 MB & 100 Mbit/s & 1 GB  \\				
			\hline
	
 \end{tabular}
 } 
 \end{table}

\begin{table*}
	\color{black}
	\centering
	\caption{Power consumption of servers in Watts}

	\label{my-label}
	\begin{tabular}{|c|c|c|c|c|c|c|c|c|c|c|c|}
		\hline
		\textbf{Servers} & \textbf{\begin{tabular}[c]{@{}c@{}}0\% \\ (sleep mode)\end{tabular}} & \textbf{10\%} & \textbf{20\%} & \textbf{30\%} & \textbf{40\%} & \textbf{\begin{tabular}[c]{@{}c@{}}50\% \\ (idle)\end{tabular}} & \textbf{60\%} & \textbf{70\%} & \textbf{80\%} & \textbf{90\%} & \textbf{\begin{tabular}[c]{@{}c@{}}100\% \\ (max)\end{tabular}} \\ \hline
		\begin{tabular}[c]{@{}c@{}}IBM x3550 M3 \\ (Interl Xeon X5670 CPU)\end{tabular} & 66 & 107 & 120 & 131 & 143 & 156 & 173 & 191 & 211 & 229 & 247 \\ \hline
		\begin{tabular}[c]{@{}c@{}}IBM x3550 M3 \\ (Intel Xeon X5675 CPU)\end{tabular} & 58.4 & 98 & 109 & 118 & 128 & 140 & 153 & 170 & 189 & 205 & 222 \\ \hline
	\end{tabular}

\end{table*}

\begin{table*}
	\caption{Parameter Configurations for Testing} 
	\newcommand{\tabincell}[2]{\begin{tabular}{@{}#1@{}}#2\end{tabular}}
	\centering 
	\footnotesize
	\begin{tabular}{|c|c|c|c|c|c|} 
		\hline 
		\textbf{Parameters} & \textbf{\tabincell{c}{P1: Optional component \\ utilization threshold}} & \textbf{\tabincell{c}{P2: Percentage of \\ optional Components}} & \textbf{\tabincell{c}{P3: Percentage of \\ connected components}} & \textbf{P4: Discount} \\
		\hline \hline
		Range & 0\% to 100\%&	0\% to 100\% &	0\% to 100\% &	0\% to 100\% \\
		\hline
		
		Categories & 25\%, 50\%, 75\%, 100\% &	25\%, 50\%, 75\%, 100\%	& 25\%, 50\%, 75\%, 100\% & varying with P1	 \\
		

		\hline 
		
		\hline 
	\end{tabular}
	\label{tab:LPer}
\end{table*}

\begin{table*}
	\caption{A Testcase Example} 
	\newcommand{\tabincell}[2]{\begin{tabular}{@{}#1@{}}#2\end{tabular}}
	\centering 
	\footnotesize
	\begin{tabular}{|c|c|c|c|c|c|} 
		\hline 
		\textbf{Testcase ID } & \textbf{\tabincell{c}{Optional component \\  utilization threshold}} & \textbf{\tabincell{c}{Percentage of \\ optional Components}} & \textbf{\tabincell{c}{Percentage of \\ connected components}}   & \textbf{Discount}\\
		\hline \hline
		TC1 & 50\%&	50\%&	25\%	& 50\%	  \\
		\hline
		

		\hline 
		
		\hline 
	\end{tabular}
	\label{tab:LPer}
\end{table*}

We use the CloudSim framework \cite{Buyya2} to simulate a cloud data center with 100 hosts. Two types of hosts and four types of VMs are modeled based on current offerings in EC2 as shown in Table 2. 
\color{black} The power models of hosts we adopted are derived from IBM System x3550 M3 with CPU Intel Xeon X5670 and X5675 \cite{SPEC}, and their power consumption at different utilization levels are demonstrated in Table 3. We assume that the idle host consumes 50\% full utilization.
\color{black}

The application modeled in CloudSim is based in a class called cloudlet. We have extended the cloudlet to model application with optional components, and each component has its corresponding CPU utilization, discount amount, and connection parameter. The components are uniformly distributed on VMs.

 We adopt the realistic workload trace from more than 1000 PlanetLab VMs \cite{PlanetLab} to create an overloaded environment \cite{AntonTPDS}.
Our experiments are simulated under one-day scheduling period and repeated 10 times based on 10 different days PlanetLab data. The brownout is invoked every 300 seconds (5 minutes per time slot) if hosts power surpasses the power threshold. The CPU resource is measured with capacity of running instructions. Assuming that the  application workload occupies 85\% resource on a VM and the VM has 1000 million instructions per second (MIPS) computation capacity, then it presents the application constantly requires 0.85 $\times$ 1000 = 850 MI to 1.0 $\times$ 1000 = 1000 MI per second in the 5 minutes. 

To reflect the impact of different configurations, we investigate a set of parameters as shown in Table 4:

1)	Optional component utilization threshold: it represents the threshold portion of utilization that is optional and can be reduced by deactivating optional components. An optional component with 25\% utilization means 25\% of application utilization is reduced if it is set as deactivated. We adjust this parameter from 0\% to 100\% and categorize it as 25\%, 50\%, 75\% and 100\%.

2)	Percentage of optional components: it represents how many components of the total components are optional. Assuming the number of all components is $num_{com}$ and the number of optional components is $num_{opt}$, then the percentage of optional components is $\frac{num_{opt}}{num_{com}}$. This parameter is varied from 0\% to 100\% and is categorized as 20\%, 50\%, 75\% and 100\%. 

3) Percentage of connected components: it represents how many components are connected among all the components. Assuming the number of connected components is $num_{connected}$, then the percentage of connected components is $\frac{ num_{connected}}{ num_{com}}$. This parameter is also varied from 0\% to 100\% and is categorized as 25\%, 50\%, 75\% and 100\%. \color{black} The connections between components are randomly generated based on  percentage of connected components.

\color{black} 4)	Discount: It represents the discount amount that allowed to be paid back to the user if components are deactivated. We assume that application maximum discount is identical to the optional component utilization threshold, for example 50\% optional component utilization threshold comes along with 50\% discount.  

\color{black} We assume that the components utilization $u(App_c)$ and discount $d(App_c)$ conform normal distribution $u(App_c) \AC N(\mu, \sigma^2), $ $d(App_c) \AC N(\mu, \sigma^2)$, the $\mu$ is the mean utilization of component utilization or discount, which is computed as the optional component utilization threshold (or discount amount) divided by the number of optional components. The $\sigma^2$ is the standard deviation of components utilization or discount. 

 Based on $\sigma^2$, we consider two component design patterns according to component utilization and discount. One pattern is that components are designed with uniform or approximate utilization and discount, which means each component is designed to require same or approximate resource amount, like there are 5 components and each component requires 10\% utilization and offers 10\% discount. We define the components as \textbf{approximate} if their utilization standard deviation and discount standard deviation are both less than 0.1. Another pattern is that components utilization and discount are conspicuous different, which means the components are designed to require quite different resource. We define the components as \textbf{different} if either their utilization standard deviation or discount standard deviation is larger than 0.1. 

\color{black} According to Table 4, Table 5 shows a testcase with configured parameters, the optional component utilization threshold is configured as 50\%, the percentage of optional utilization is configured as 50\%, the percentage of connected components is set as 25\% and the discount is 50\%.


Table 6 demonstrates an application components example fits the configurations in Table 5. This application consists of 8 components: 4 of them (50\%) are optional components.  Each component has utilization, discount and connected relationship with other components: the optional component utilization threshold is 50\% (the utilization sum of component 5, 6, 7 and 8), there are 2 components (20\%) of all components are connected (component 5 and 6) and the total discount of optional components is 50\%.

\subsection{Results and Analysis}\
\color{black}In this section, we compare EEBA performance with two baselines algorithms:

\color{black} 1)\textbf{ VM Placement and Consolidation  algorithm (PCO)}: the algorithm is described in Section 4.1. 
\color{black} We configure its upper threshold as 0.8 and the lower threshold as 0.2.

\begin{table}[!ht]
	\caption{An Application Component Example1} 
	\newcommand{\tabincell}[2]{\begin{tabular}{@{}#1@{}}#2\end{tabular}}
	\resizebox{0.48\textwidth}{!}{%
		\centering 
		\footnotesize
		\begin{tabular}{|l|l|l|l|l|} 
			\hline  \tabincell{c}{Components\\ ID} & \tabincell{c}{Mandatory /\\ Optional} & Utilization & Discount  &  \tabincell{c}{Connected}  \\ 
			\hline
			\hline Comp 1 & Mandatory & 10\% & 10\%  & N/A  \\ 
			\hline Comp 2 & Mandatory & 10\% & 10\%  & N/A  \\
			\hline Comp 3 & Mandatory & 20\% & 20\% & N/A  \\
			\hline Comp 4 & Mandatory & 10\% & 10\% & N/A  \\				
			\hline Comp 5 & Optional & 5\% & 5\% & Comp8  \\
			\hline Comp 6 & Optional & 10\% & 10\% & Comp7  \\
			\hline Comp 7 & Optional & 15\% & 20\% &  N/A \\
			\hline Comp 8 & Optional & 20\% & 15\% & N/A  \\
			\hline
			

			\hline 
			
		\end{tabular}
	}
\end{table}

\color{black} 2) \textbf{Utilization-based Probabilistic VM consolidation algorithm (UBP)} \cite{Chen}: in the VM placement, UBP adopts the same approach as PCO: sorting all the VMs in decreasing order based on their utilization and allocating each VM to the host that increases the least power consumption.  In the VM consolidation phase, UBP applies a probabilistic method \cite{Mastroianni} to select VMs from overloaded host. The probabilistic method calculates the migration probability based on PM utilization $u$ as :

  {\small     \begin{equation}
  	f_m(u) = (1 - \frac{u-1}{1- T_h})^\lambda
  	\end{equation} }
where $f_m(u)$ is the migration probability, $T_h$ is the upper threshold for detecting overloads and $\lambda $ is a constant to adjust probability. 
\color{black} We configure the $T_h = 0.8$ and $\lambda = 1$. 

In EEBA, we also configure $TP = 0.8$ that is as same as the upper threshold in PCO and $T_h$ in UBP.

 \color{black} We separately conduct experiments for the two design patterns to evaluate algorithm performance. With approximate components, our proposed policies LUFCS, LPFCS and HUPRFCS select the same components, so we focus on comparing PCO, UBP, NUFCS and LUFCS policies, which represent baseline algorithms without brownout, EEBA with single component selection policy and EEBA with multiple components selection policy respectively. While with different components, we focus on comparing the LUFCS, LPFCS and HUPRFCS policies to evaluate performance of different multiple components selection policies.      

In addition, as introduced in Section 3.1, components may be connected. Therefore, to investigate the effects of individual component selection and connected components selection, we separately run experiments for components without connections and connected components.

As PCO and UBP performance are not influenced by the parameters in Table 4, we firstly obtain their results as baselines. The PCO leads to 345.3 kWh with 95\% confidence interval (CI): (336.9, 353.7), and UBP reduces this value to 328.5kWh with 95\% CI: (321.1, 335.9).  Both these two algorithms offer no discount and no disabled utilization. Because UBP performance is better than PCO, we set PCO as the benchmark. Referring to equation (9), $E_b$ is set as 345.3, so PCO algorithm efficiency $Eff_{PCO} = 345.3 / 345.3 + 0.0 = 1.0$; for UBP algorithm, its efficiency is $Eff_{UBP} = 321.1 / 345.3 + 0.0 = 0.95$.

\subsubsection{Components without Connections}

 \begin{figure*}[ht]
 	\centering
 	\includegraphics[width=1.0\linewidth]{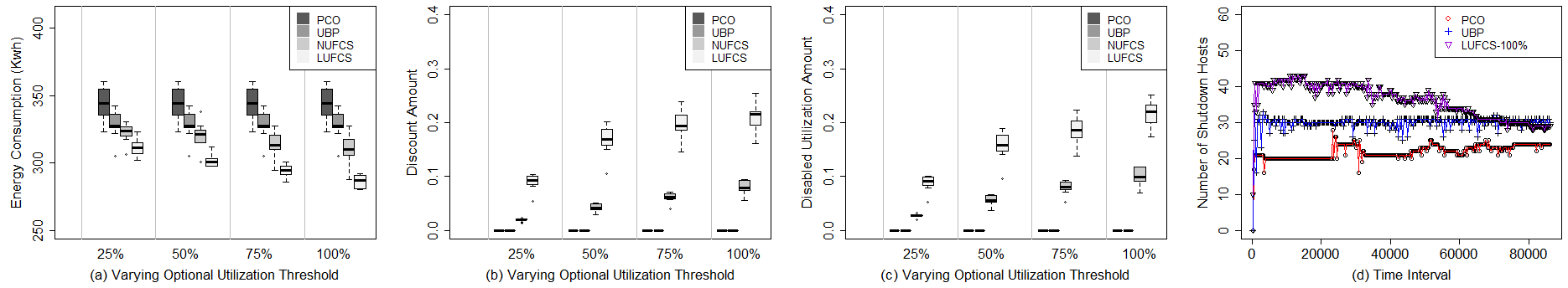}
 	\caption[VarPerOptCom]{Comparison by Varying Optional Utilization Threshold for Approximate Components}
 	
 	\label{fig:VarPerOptCom}
 \end{figure*}

 \begin{figure*}[ht]
 	\centering
 	\includegraphics[width=1.0\linewidth]{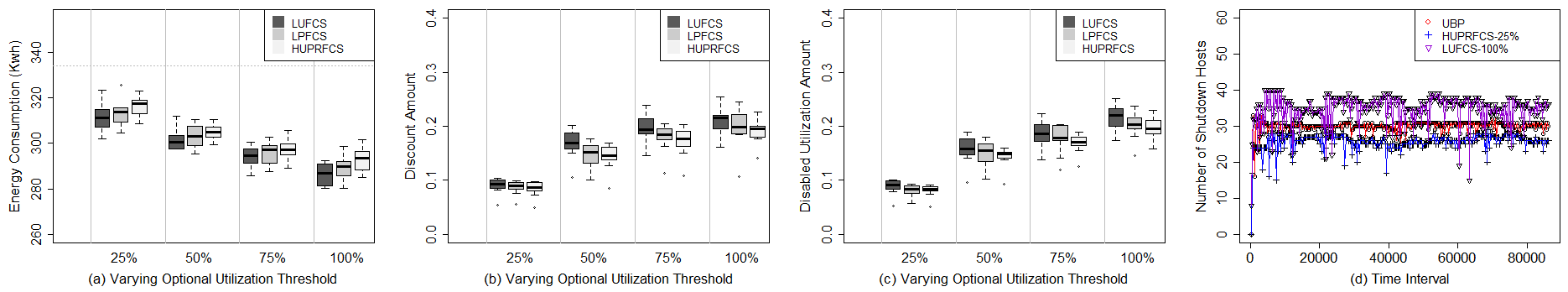}
 	\caption[VarConnectedCom]{Comparison by Varying Optional Utilization Threshold for Different Components}
 	
 	\label{fig:VarConnectedCom}
 \end{figure*}

 \begin{figure*}[ht]
 	\centering
 	\includegraphics[width=1.0\linewidth]{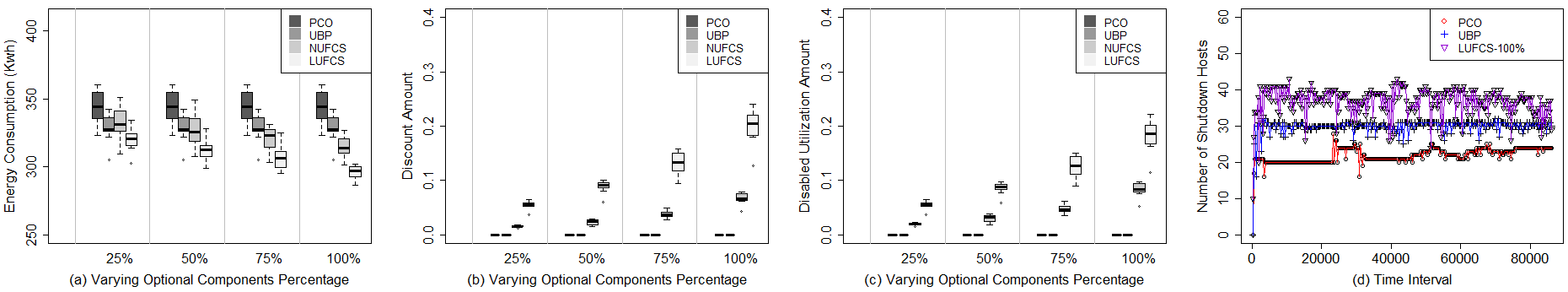}
 	\caption[VarOptComUtil]{Comparison by Varying Optional Component Percentage for Approximate Components}
 	
 	\label{fig:VarOptComUtil}
 \end{figure*}

  \begin{figure*}[ht]
  	\centering
  	\includegraphics[width=1.0\linewidth]{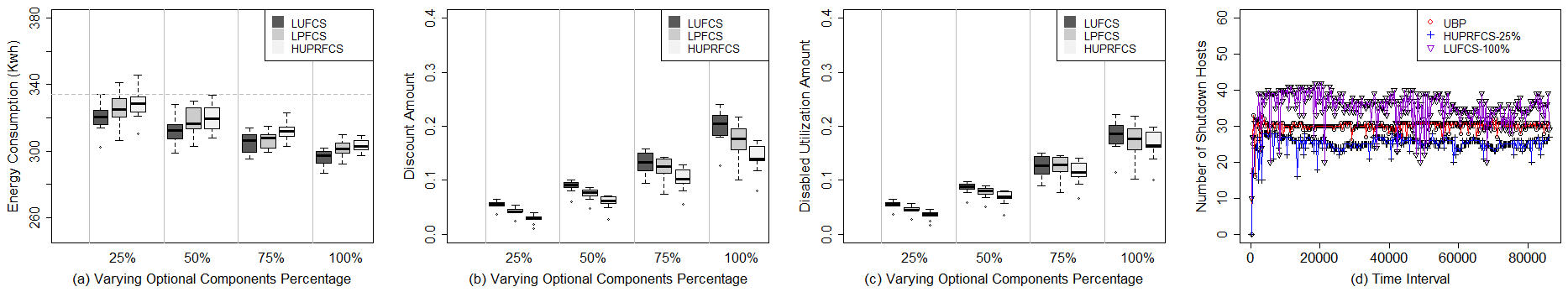}
  	\caption[VarComPrices]{Comparison by Varying Optional Component Percentage for Different Components}
  	
  	\label{fig:VarCompoPrices}
  \end{figure*}

  \begin{figure*}[ht]
  	\centering
  	\includegraphics[width=1.0\linewidth]{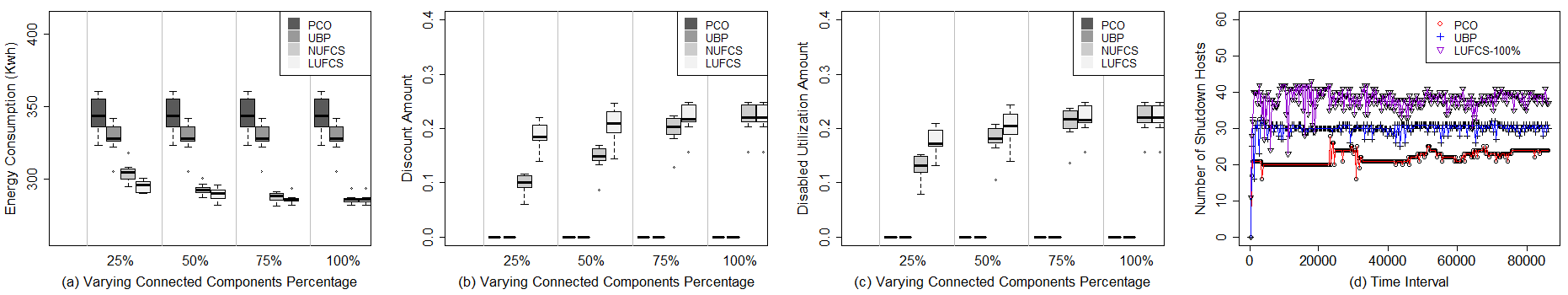}
  	\caption[VarCompoPower]{Comparison by Varying Percentage of Connected Components for Approximate Components}
  	
  	\label{fig:VarPowerThreshold}
  \end{figure*}

  \begin{figure*}[ht]
  	\centering
  	\includegraphics[width=1.0\linewidth]{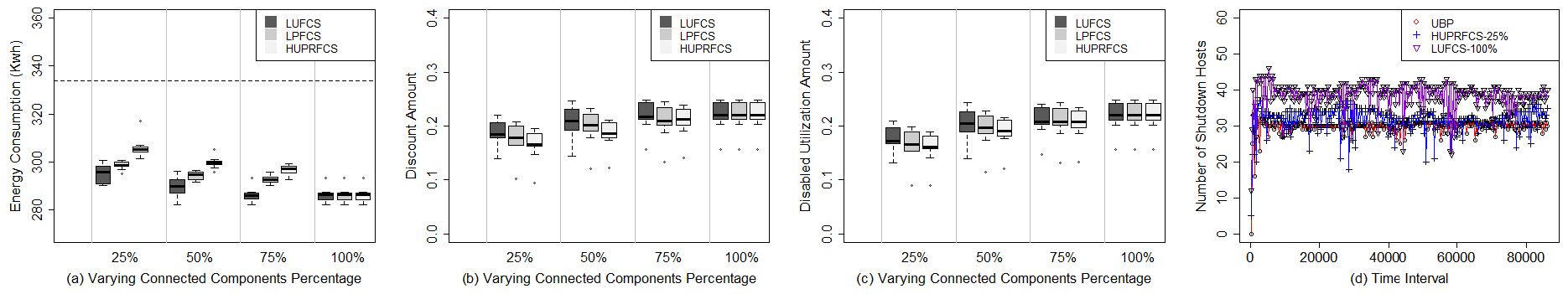}
  	\caption[VarCompoPower]{Comparison by Varying Percentage of Connected Components for Different Components}
  	
  	\label{fig:VarPowerThreshold}
  \end{figure*}
  
    \begin{figure*}[ht]
    	\centering
    	\includegraphics[width=1.0\linewidth]{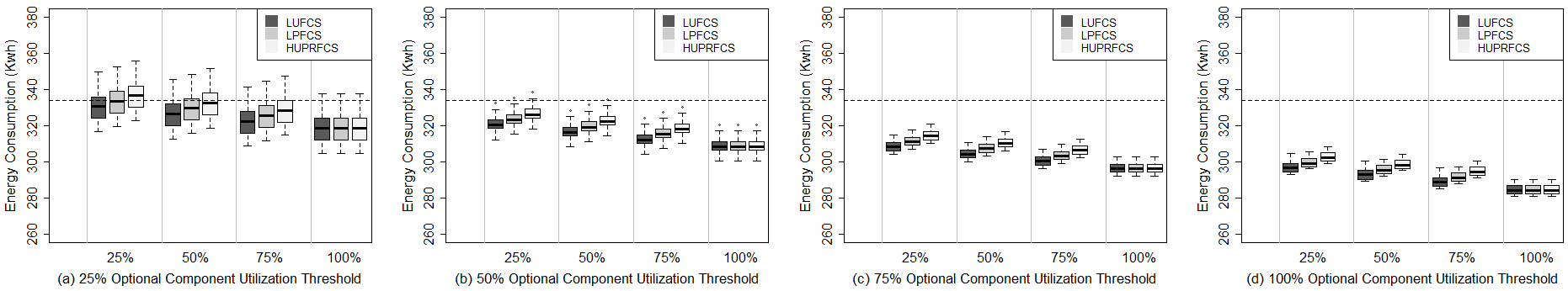}
    	\caption[VarCompoPower]{Energy Consumption Comparison by Varying Percentage of Connected Components and Optional Component Utilization Threshold}
    	
    	\label{fig:VarPowerThreshold}
    \end{figure*}

     \begin{figure*}[ht]
     	\centering
     	\includegraphics[width=1.0\linewidth]{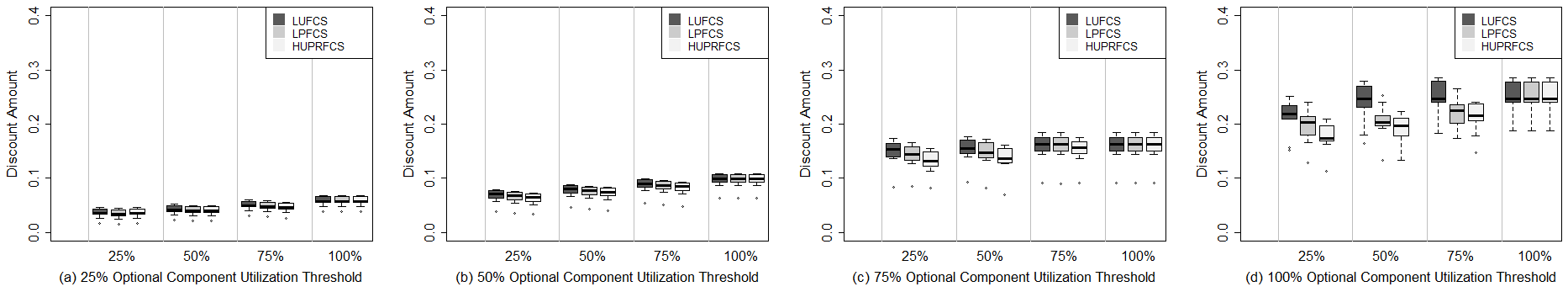}
     	\caption[VarCompoPower]{Discount Amount Comparison by Varying Percentage of Connected Components  and Optional Component Utilization Threshold}
     	
     	\label{fig:VarPowerThreshold}
     \end{figure*}

       \begin{figure*}[ht]
       	\centering
       	\includegraphics[width=1.0\linewidth]{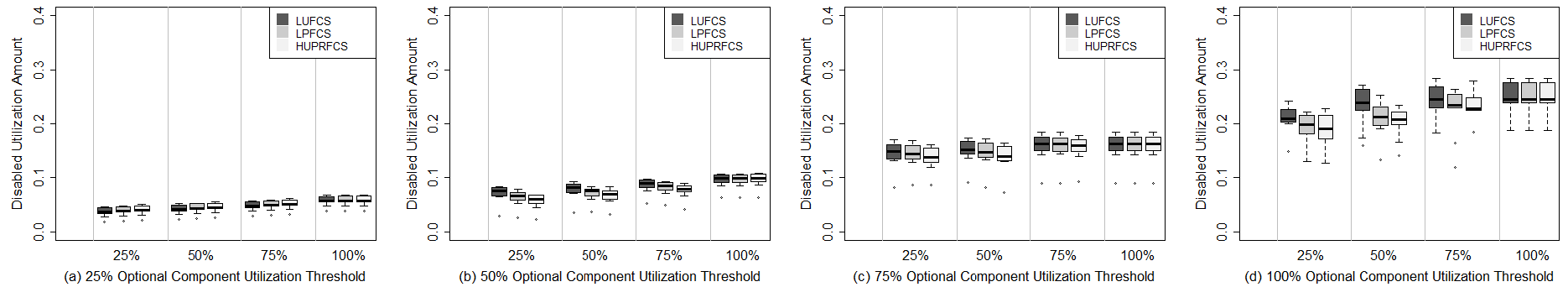}
       	\caption[VarCompoPower]{Disabled Utilization Amount Comparison by Varying Percentage of Connected Components and Optional Component Utilization Threshold}
       	
       	\label{fig:VarPowerThreshold}
       \end{figure*}

\textit{\textbf{1) Varying Optional Component Utilization Threshold}}

Fig. 2 shows the comparison between PCO, UBP, NUFCS and LUFCS when components are approximate by varying the optional utilization threshold (the percentage of optional components is fixed as 50\%). Fig. 2a shows the energy consumption of these policies respectively. NUFCS and LUFCS can save more energy when the optional component utilization threshold is larger. However, more discount amount is also offered to users according to Fig. 2b. The reason lies in that Fig 2c and Fig 2d demonstrate that more utilization amount is disabled in NUFCS and LUFCS, and more hosts are shutdown by these policies, which contributes to more energy reduction. We use UBP and NUFCS with 100\% optional utilization threshold to compare the number of shutdown hosts with PCO, which shows the maximum and minimum number of shutdown hosts in this series of experiments. The number of shutdown hosts of other experiments falls between the UBP and LUFCS-100\% lines in Fig 2d. Compared with UBP, NUFCS reduces 2\% to 7\% energy and LUFCS reduces 6\% to 21\% energy while 1\% to 6\% and 8\% to 22\% discount amount are offered respectively. 

Fig. 3 shows LUFCS, LPFCS and HUPRFCS policies' effects on energy and discount amount when components are different and optional utilization threshold increases, more energy is reduced and more discount amount is offered. As Fig. 3a and Fig. 2a illustrate, when components are different, LUFCS cannot save as much energy as when components are approximate. For example, LUFCS-100\% in Fig 3a shows it reduces maximum 15\% energy (the dotted line represents UBP energy consumption), while LUFCS-100\% in Fig. 2a saves 21\% energy. Therefore, our proposed policies work better when components are designed with approximate resource requirement, which also shows the value of proper design of components or microservices.  According to Fig. 3a and Fig. 3b, LUFCS reduces the maximum energy, but also offers the maximum discount amount, which gives LUFCS $<$ LPFCS $<$ HUPRFCS in energy and LUFCS $>$ LPFCS $>$ HUPRFCS in discount amount. 
We conduct paired t-tests for energy consumption of these policies, when optional utilization threshold is 25\%, the\textit{ p-value} for LUFCS-25\% and HUPRFCS-25\% is 0.12, which shows nonstatistically significant differences between these policies, while the optional utilization threshold increases, the \textit{p-value} for LUFCS-50\%  and LPFCS-50\% is 0.038 and the \textit{p-value} for LPFCS-50\% and HUPRFCS-50\% is 0.046, which shows there are statistically significant differences in energy consumption. As shown in Fig. 3c and Fig. 3d, it reflects more utilization amount is disabled and more hosts are shutdown in LUFCS as shown. The different effects between these policies come from the LUFCS selects components without considering discount, as it can select as many components as possible until achieving expected utilization reduction. While other two policies consider discount amount and do not deactivate as many components as in LUFCS.     

\textit{\textbf{2) Varying Percentage of Optional Components}}

Fig 4 shows the results when components are approximate by varying the percentage of optional components (the optional component utilization threshold is fixed as 50\%). Fig. 4a and Fig. 4b illustrate that in comparison to PCO and UBP, more energy is saved and more discount amount is offered with the increase of the optional components in NUFCS and LUFCS, which results from more options of components are available to be selected. In comparison to UBP, when more than 25\% components are optional, NUFCS saves 1\% to 7\% energy and offers maximum 12\% discount amount, and LUFCS saves 5\% to 19\% energy but offers 5\% to 20\% discount amount. As shown in Fig. 4c and Fig. 4d, compared with UBP, LUFCS disables maximum 19\% utilization amount and more than 8 hosts averagely. 

Fig. 5 compares LUFCS, LPFCS and HUPRFCS policies for different components when varying the percentage of optional components. The results in Fig. 5a and Fig 5b show that these policies save more energy when optional components increases, and show LUFCS $<$ LPFCS $<$ HUPRFCS in energy as well as LUFCS $>$ LPFCS $>$ HUPRFCS in discount amount. As demonstrated in Fig. 5c and Fig. 5d, LUFCS disables more utilization amount than other two policies and shuts down the maximum number of hosts when with 100\% optional components. Through paired t-tests for 25\% optional components, we observe the p-value for LUFCS and LPFCS is 0.035, which shows statistically significant differences. But the p-value for LPFCS and HUPRFCS is 0.095, which shows nonstatistically significant different. Similar p-values are also observed when more optional components are provided. 

Although LUFCS with 100\% optional components saves about 19\% and 16\% energy for approximate components and different components respectively, it is not recommended to set all components as optional since too much discount amount is offered. 
We will discuss policies selection considering the trade-offs in the Section 5.2.3.

\subsubsection{Connected Components}

\begin{table*}[]
	\centering
	\caption{Recommended Policies for Components without Connections under Different Configurations}
	\label{my-label}
	\resizebox{\textwidth}{!}{%
	\begin{tabular}{|c|c|c|c|c|c|}
		\hline
		\begin{tabular}[c]{@{}c@{}}Components \\ Design Pattern\end{tabular}               & \begin{tabular}[c]{@{}c@{}}Discount \\ Constraint\end{tabular} & \begin{tabular}[c]{@{}c@{}}Optional Component \\ Utilization Threshold\end{tabular} & \begin{tabular}[c]{@{}c@{}}Percentage of \\ Optional Components\end{tabular} & Recommend Policy & Algorithm Efficiency                                    \\ \hline
		\multirow{2}{*}{\begin{tabular}[c]{@{}c@{}}Approximate \\ Components\end{tabular}} & $\leq$ 5\%                                                      & \multirow{2}{*}{$\leq$ 100\%}                                                         & \multirow{2}{*}{$\leq$ 100\%}                                                  & NUFCS            & 0.947 with 95\% CI: (0.941, 0.953)                       \\ \cline{2-2} \cline{5-6} 
		& \textgreater 5\%                                              &                                                                                     &                                                                              & LUFCS            & 0.91 with 95\% CI: (0.887, 0.933)                        \\ \hline
		\multirow{3}{*}{\begin{tabular}[c]{@{}c@{}}Different \\ Components\end{tabular}}                                              & \multirow{3}{*}{$\leq$ 100\%}                                    & $\leq$ 50\%                                                                           & \multirow{3}{*}{$\leq$ 100\%}                                                  & LUFCS            & 0.918 with 95\% CI: (0.895, 0.941)                       \\ \cline{3-3} \cline{5-6} 
		&                                                                & 50\%-75\%                                                                           &                                                                              & LPFCS            & 0.913 with 95\% CI: (0.891, 0.936)                      \\ \cline{3-3} \cline{5-6} 
		&                                                                & \textgreater 75\%                                                                   &                                                                              & HUPRFCS          & \multicolumn{1}{l|}{0.908 with 95\% CI: (0.885, 0.931)} \\ \hline
	\end{tabular}
}
\end{table*}

\begin{table*}[]
	\centering
	\caption{Recommended Policies for Connected Components under Different Configurations}
\label{my-label}
\resizebox{\textwidth}{!}{%
	\begin{tabular}{|c|c|c|c|c|c|c|}
		\hline
		\begin{tabular}[c]{@{}c@{}}Components \\ Design Pattern\end{tabular}               & \begin{tabular}[c]{@{}c@{}}Discount \\ Constraint\end{tabular} & \begin{tabular}[c]{@{}c@{}}Percentage of\\ Connected Components\end{tabular} & \begin{tabular}[c]{@{}c@{}}Optional Component \\ Utilization Threshold\end{tabular} & \begin{tabular}[c]{@{}c@{}}Percentage of \\ Optional Components\end{tabular} & \begin{tabular}[c]{@{}c@{}}Recommended\\  Policy\end{tabular} & \begin{tabular}[c]{@{}c@{}}Algorithm Efficiency\\ with 95\% CI\end{tabular} \\ \hline
\multirow{2}{*}{\begin{tabular}[c]{@{}c@{}}Approximate \\ Components\end{tabular}} & $\leq$ 5\%                                                    & \multirow{2}{*}{$\leq$ 100\%}                                                & $\leq$ 100\%                                                                        & \multirow{2}{*}{$\leq$ 100\%}                                                & NUFCS                                                         & 0.901 (0.852, 0.953)                                                        \\ \cline{2-2} \cline{4-4} \cline{6-7} 

		& \textgreater 5\%                                              &                                                                              & $\leq$ 100\%                                                                        &                                                                              & LUFCS                                                         & 0.877 (0.852, 0.902)                                                        \\ \hline
		\multirow{5}{*}{\begin{tabular}[c]{@{}c@{}}Different \\ Components\end{tabular}}   & \multirow{5}{*}{$\leq$ 100\%}                                  & \multirow{2}{*}{$\leq$ 50\%}                                                 & $\leq$ 50\%                                                                         & \multirow{5}{*}{$\leq$ 100\%}                                                & LUFCS                                                         & 0.895 (0.859, 0.931)                                                        \\ \cline{4-4} \cline{6-7} 
		&                                                                &                                                                              & \textgreater 50\%                                                                   &                                                                              & LPFCS                                                         & 0.89 (0.858, 0.922)                                                         \\ \cline{3-4} \cline{6-7} 
		&                                                                & \multirow{2}{*}{50\%-75\%}                                                   & $\leq$ 50\%                                                                         &                                                                              & LPFCS                                                         & 0.886 (0.855, 0.917)                                                        \\ \cline{4-4} \cline{6-7} 
		&                                                                &                                                                              & \textgreater 50\%                                                                   &                                                                              & HUPRFCS                                                       & 0.881 (0.856, 0.906)                                                        \\ \cline{3-4} \cline{6-7} 
		&                                                                & \textgreater 75\%                                                            & $\leq$ 100\%                                                                        &                                                                              & HUPRFCS                                                       & 0.880 (0.85, 0.91)                                                          \\ \hline
	\end{tabular}%
}
\end{table*}

After investigating the components without connections, we move to investigate connected components.  As mentioned in Algorithm 2, in these cases, our proposed policies treat the connected components together and use their average utilization or discount to sort. Fig. 6 shows the PCO, UBP, NUFCS and LUFCS for approximate components when varying the percentage of connected components (optional component utilization threshold and percentage of optional components are both fixed as 50\%). Fig. 6a shows that the connected components affects the NUFCS impressively. The energy consumption drops heavily in NUFCS when the percentage of connected components increases, i.e., from 9\% to 21\% reduction compared with UBP. While in LUFCS, the connected components do not affect its performance significantly. Although the energy consumption is also reduced when the percentage of connected components increases, energy consumption drops slowly from 14\% to 21\%. When 100\% components are connected, NUFCS and LUFCS produce the same effects. As shown in Fig. 6b, with the increase of connected components, discount amount increases fast from 10\% to 23\% in NUFCS while slowly in LUFCS from 17\% to 23\%. NUFCS and LUFCS both offer same discount amount when all the components are connected. 
For the cases that save more energy, like NUFCS or LUFCS with 100\% connected components, Fig. 6c and Fig. 6d show that more utilization amount is disabled and more hosts are shutdown than baseline algorithms.

Fig .7 illustrates the comparison of LUFCS, LPFCS and HUPRFCS for different components when varying the percentage of connected components. Fig. 7a shows that when connected components are larger than 75\%, these policies do not result in significant differences, this is due to when the percentage of connected components increases, similar deactivated component lists are obtained although these components may be put into the list in different orders by these policies. Apparent differences for discount amount and disabled utilization amount are illustrated in Fig. 7b and Fig. 7c when connected components are less than 75\%, like LUFCS reduces 2\% to 5\% energy than LPFCS and 5\% to 10\% energy than HUPRFCS, LUFCS offers 4\% to 10\% more discount than LPFCS and 9\% to 15\% more discount amount than HUPRFCS.  Fig. 7d shows that when components are connected, more hosts are shutdown than UBP.  In summary, our proposed multiple components selection policies works better under components with lower connected percentage up to 75\%, which enables to provide multiple choices for service provider rather than providing same effects.

To evaluate the effects of combined parameters, we vary optional component utilization threshold and percentage of connected components together. We choose the optional component utilization threshold, as this parameter shows more significant effects than the percentage of optional component in energy and discount. Fig. 8 to 10 demonstrate the energy consumption, discount amount and disabled utilization amount separately when varying these two parameters together. Each subfigure is with fixed optional component utilization threshold and variable percentage of connected components, for example, Fig. 8a represents energy consumption when optional component utilization threshold is 25\% and percentage of connected component is varied from 25\% to 100\%. 
Fig. 8 shows that energy is reduced when connected components increases or larger optional component utilization threshold is given. 
For the compared policies, LUFCS, LPFCS and HUPRFCS show similar results when optional component utilization threshold is below 25\% or percentage of connected components is above 75\%. This is  because when optional component utilization threshold is low, the disabled utilization is quite close for different policies, and higher percentage of connected components also contributes to deactivating the same list of components. For other cases that show statistically significant differences in energy consumption with \textit{p-value} less than 0.05, like in Fig. 7(a), the results are given as LUFCS $\leq$ LPFCS $\leq$ HUPRFCS. In these cases, Fig. 9 and Fig. 10 also show that LUFCS $>$ LPFCS $>$ HUPRFCS in discount amount and disabled utilization.

In conclusion, EEBA algorithm saves more energy than the VM consolidation approaches without brownout, like PCO and UBP. It is noticed that our experiments are mainly focused on optimizing servers energy consumption, so the network infrastructure energy consumption is not optimized. Since the component selection policies in brownout controller can be modelled into applications, like in \cite{Durango}, they are insensitive to network infrastructures. 

  \subsubsection{Policy Selection Recommendation}

To help make choices for component selection policies, we use the equation (9) to calculate their algorithm efficiency and summarize suitable policies under different configurations to achieve better energy efficiency. We consider energy consumption and discount with the same importance, so the $\alpha$ is set as 1. Table 7 shows the results for components without connections and Table 8 presents the results for the connected components.  

To sum up, for components without connections, 1) when the components are approximate, NUFCS fits in the configurations when service provider allows maximum 5\% discount and LUFCS performs better when more discount amount is allowed by service provider. 2) When the components are different, although the discount constraint is not as important as in the approximate components cases, the policies are picked out by other parameters, for instance, LUFCS achieves the best efficiency with less than 50\% optional component utilization threshold, LPFCS overwhelms others with 50\% to 75\% optional component utilization threshold, HUPRFCS performs the best efficiency with more than 75\% optional components utilization threshold.

For connected components, the suitable conditions are more complex: 1) when the components are approximate, NUFCS is recommended if discount amount is limited under 5\% and LUFCS is suggested if more than 5\% discount amount is allowed; 2) when the components are different, recommended policy changes via different configurations. For example, when connected components are less than 50\%, if optional component utilization threshold is less than 50\%, LUFCS is recommended; if optional component utilization threshold is larger than 50\%, LPFCS is recommended. 
When connected components are between 50\% and 75, LPFCS is recommended for optional component utilization threshold that is not larger than 50\%, HUPRFCS is recommended for optional component utilization threshold larger than 50\%. When more than 75\% components are connected, any policy achieves quite close results, HUPRFCS is a choice.

\section{Conclusions and Future Work}

Brownout has been proven effective to solve the overloaded situation in cloud data centers. Additionally, brownout can also be applied to reduce energy consumption. We introduce the brownout enabled system model by considering application components, which are either mandatory or optional. In the model, the brownout controller can deactivate the optional components to reduce data center energy consumption while offering discount to users. We also propose a brownout enabled algorithm to determine when to use brownout and how much utilization on a host is reduced. Then we present a number of policies to select components and investigate their effects on energy consumption and discount offering. 

In our experiments, we consider different configurations, such as components without connections, connected components, approximate components, different components and etc.
The results show that these proposed policies save more energy than the baselines PCO and UBP. The comparison of proposed policies demonstrates that these policies fit in different configurations. Considering the discount amount offered by a service provider, NUFCS is recommended when a small amount of discount (like less than 5\%) is offered, as it can reduce maximum 7\% energy consumption in contrast to UBP. When more discount amount (like more than 5\%) is allowed by service provider, other multiple components selection policies are recommended, for example, compared with UBP, HUPRFCS saves more than 20\% energy with 10\% to 15\% discount amount.

As for future work, to avoid ineffective deactivation, we plan to investigate Markov Decision Process (MDP) to determine whether the energy consumption would be reduced if some components are deactivated. 
We also plan to implement proposed policies and deploy them to OpenStack 
and web application system such as Apache web server.


%



\ifCLASSOPTIONcompsoc
  \section*{Acknowledgments}
\else
  \section*{Acknowledgment}
\fi

This work is supported by China Scholarship Council (CSC), Australia Research Council Future Fellowship and Discovery Project Grants.

\ifCLASSOPTIONcaptionsoff
  \newpage
\fi



%

\bibliographystyle{IEEEtran}
\bibliography{tsuc}

%


%

\begin{IEEEbiography}
	[{\includegraphics[width=1in,height=1.25in,clip,keepaspectratio]{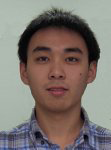}}]{Minxian Xu}
	received the BSc degree
	in 2012 and the MSc degree in 2015, both in
	software engineering
	from University of Electronic Science and Technology of China.
	He is working towards the PhD degree at the
	Cloud Computing and Distributed Systems
	(CLOUDS) Laboratory, Department of Computing
	and Information Systems, the University of
	Melbourne, Australia. His research interests include resource scheduling and optimization in cloud computing.
\end{IEEEbiography}
\begin{IEEEbiography}
	[{\includegraphics[width=1in,height=1.25in,clip,keepaspectratio]{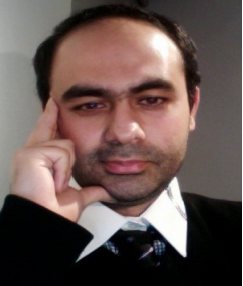}}]{Amir Vahid Dastjerdi}
	is a Research Fellow
	with the Cloud Computing and Distributed
	Systems (CLOUDS) Laboratory at the University
	of Melbourne. His current research
	interests include Cloud service coordination,
	scheduling, and resource provisioning using
	optimization, machine learning, and artificial
	intelligence techniques.

\end{IEEEbiography}
\begin{IEEEbiography}
	[{\includegraphics[width=1in,height=1.25in,clip,keepaspectratio]{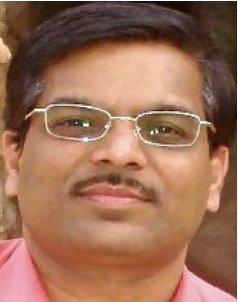}}]{Rajkumar Buyya}
is professor and future fellow of
the Australian Research Council, and the director
of the Cloud Computing and Distributed Systems
(CLOUDS) Laboratory at the University of
Melbourne, Australia. He is also serving as the
founding CEO of Manjrasoft, a spin-off company
of the University, commercializing its innovations
in Cloud Computing. He has authored more than
425 publications and four text books including
“Mastering Cloud Computing” published by
McGraw Hill and Elsevier/Morgan Kaufmann,
2013 for Indian and international markets, respectively. He is one of the
highly cited authors in computer science and software engineering
worldwide. 
Microsoft Academic Search Index ranked him as the world’s
top author in distributed and parallel computing between 2007 and 2012.
Software technologies for grid and cloud computing developed under his
leadership have gained rapid acceptance and are in use at several academic institutions and commercial enterprises in 40 countries around
the world. 
He has led the establishment and development of key community
activities, including serving as foundation chair of the IEEE Technical
Committee on Scalable Computing and five IEEE/ACM
conferences. 
His these contributions and international research leadership
are recognized through the award of “2009 IEEE Medal for Excellence
in Scalable Computing” from the IEEE computer society.

\end{IEEEbiography}





\end{document}